\documentclass[12pt]{article}
\usepackage{color}
\usepackage{amssymb,amsmath}
\usepackage[dvips]{graphicx}
\newcommand{\bea}{\begin{eqnarray}}
\newcommand{\eea}{\end{eqnarray}}
\begin{document}
\setlength{\baselineskip}{0.7cm}
\setlength{\baselineskip}{0.7cm}
\begin{titlepage} 
\begin{center}{\Large\bf Diphoton Decay Excess and 125 GeV Higgs Boson \\
\vspace*{2mm}
in Gauge-Higgs Unification}
\end{center}
\vspace*{10mm}
\begin{center}
{\large Nobuhito Maru}$^{a}$ 
and 
{\large Nobuchika Okada}$^{b}$
\end{center}
\vspace*{0.2cm}
\begin{center}
${}^{a}${\it Department of Physics, and
Research and Education Center for Natural Sciences, \\
Keio University, Hiyoshi, Yokohama 223-8521, Japan}
\\[0.2cm]
${}^{b}${\it Department of Physics and Astronomy, University of Alabama, \\
Tuscaloosa, Alabama 35487, USA} 
\end{center}
\vspace*{2cm}
\begin{abstract} 

In the context of gauge-Higgs unification scenario  
 in a 5-dimensional flat spacetime, 
 we investigate Higgs boson production via gluon fusion 
 and its diphoton decay mode at the LHC. 
We show that the signal strength of the Higgs diphoton decay mode 
 observed at ATLAS and CMS, which is considerably larger 
 than the Standard Model expectation, 
 can be explained by a simple gauge-Higgs unification model 
 with color-singlet bulk fermions to which  
 a half-periodic boundary condition is assigned. 
The bulk fermions also play a crucial role 
 in reproducing the observed Higgs boson mass around 125 GeV. 

\end{abstract}
\end{titlepage}

\section{Introduction}

It was recently announced by ATLAS \cite{ATLAS} and CMS \cite{CMS} 
 collaborations that Higgs(-like) boson was discovered 
 at the Large Hadron Collider (LHC). 
Although the observed data for a variety of Higgs boson 
 decay modes are found to be consistent 
 with the Standard Model (SM) expectations, 
 the diphoton decay mode shows the signal strength 
 considerably larger than the SM prediction. 
Since the Higgs-to-diphoton coupling arises at the quantum level 
 even in the SM, there is a good chance that the deviation 
 originates from a certain new physics effect. 
This has motivated many recent studies for explanation 
 of the deviation in various extensions of the SM 
 with supersymmetry~\cite{AHC-SUSY} or 
 without supersymmetry~\cite{AHC-NonSUSY}.

In this paper, we investigate Higgs production 
 via the gluon fusion and its diphoton decay 
 in gauge-Higgs unification (GHU) \cite{GH}. 
The GHU scenario offers a solution to the gauge 
 hierarchy problem without invoking supersymmetry, 
 where the SM Higgs doublet is identified 
 as an extra spatial component of the gauge field 
 in higher dimensional theory. 
The scenario predicts various finite physical observables 
 such as Higgs potential~\cite{1loopmass, 2loop}, 
 $H\to gg, \gamma\gamma$~\cite{MO, Maru}, 
 the anomalous magnetic moment $g-2$~\cite{g-2}, 
 and the electric dipole moment~\cite{EDM}, 
 thanks to the higher dimensional gauge symmetry, 
 irrespective of the non-renormalizable theory.

In our previous paper \cite{MO} based on a simple GHU model, 
 we have calculated loop-contributions of Kaluza-Klein (KK) 
 modes to the Higgs-to-digluon and Higgs-to-diphoton couplings 
 and found that the KK mode contributions are destructive 
 to the SM contributions by corresponding SM particles. 
This is a remarkable feature of the GHU, 
 closely related to the absence of the quadratic divergence 
 in Higgs self-energy corrections. 
In this paper we revisit this analysis 
 for a simple extension of our previous GHU model. 
Introducing color-singlet bulk fermions with a half-periodic 
 boundary condition, we investigate their effects  
 on the Higgs-to-diphoton coupling and Higgs boson mass. 
We will show that the bulk fermions of certain representations 
 of the bulk gauge group can easily enhance 
 the Higgs-to-diphoton coupling with appropriately chosen 
 electric charges. 
The bulk fermions also play a crucial role 
 to achieve a Higgs boson mass around 125 GeV. 
Without the bulk fermions, Higgs boson mass is predicted 
 to be too small, less than 100 GeV.

The plan of this paper is as follows. 
In the next section, we consider a 5-dimensional GHU model 
 based on the gauge group $SU(3)\times U(1)'$ 
 with an orbifold $S^1/Z_2$ compactification \cite{SSS, CCP}. 
As simple explicit examples, 
 we introduce color-singlet bulk fermions 
 of ${\bf 10}$ and ${\bf 15}$ representations 
 and impose a half-periodic boundary condition for them. 
$U(1)'$ charges for the bulk fermions are appropriately assigned. 
In this context, we calculate Higgs production via gluon fusion 
 and diphoton decay processes.
The KK modes of the bulk fermions contribute to 
 the Higgs-to-diphoton coupling constructively to 
 the $W$-boson loop corrections in the SM 
 and enhance the Higgs diphoton  branching ratio. 
The magnitude of enhancement is determined by 
 $U(1)'$ charges and the representation of the bulk fermion, 
 and a suitable choice of them can account for the observed 
 signal strength of Higgs diphoton decay mode. 
In section 3, we estimate Higgs boson mass 
 using a 4-dimensional effective theory approach developed 
 in Ref.~\cite{GHcondition}, where Higgs boson mass 
 is determined via 1-loop renormalization group equation (RGE) 
 of the Higgs quartic coupling 
 with the ``gauge-Higgs condition" \cite{GHcondition}. 
We find that the introduced bulk fermions play a 
 crucial role to achieve a Higgs boson mass around 125 GeV 
 through the RGE running of the Higgs quartic coupling. 
Section 4 is devoted for the conclusions.

\section{Higgs production and diphoton decay in GHU}
We consider a GHU model based on the gauge group 
 $SU(3) \times U(1)'$ in a 5-dimensional flat space-time 
 with orbifolding on $S^1/Z_2$ with radius $R$ of $S^1$. 
In our setup of bulk fermions, we follow Ref.~\cite{CCP}: 
 the up-type quarks except for the top quark, 
 the down-type quarks and the leptons are embedded 
 respectively into ${\bf 3}$, $\overline{{\bf 6}}$, 
 and ${\bf 10}$ representations of $SU(3)$. 
In order to realize the large top Yukawa coupling, 
 the top quark is embedded into a rank $4$ representation 
 of $SU(3)$, namely $\overline{{\bf 15}}$. 
The extra $U(1)'$ symmetry works to yield 
 the correct Weinberg angle, and the SM $U(1)_Y$ gauge boson 
 is given by a linear combination between the gauge bosons 
 of the $U(1)'$ and the $U(1)$ subgroup in $SU(3)$ \cite{SSS}\footnote{
 It is known that the correct Weinberg angle can also be obtained 
 by introducing brane localized gauge kinetic terms \cite{SSS}, 
 but we do not take this approach in this paper.}.  
We assign appropriate $U(1)'$ charges for bulk fermions 
 to give the correct hyper-charges for the SM fermions.

The boundary conditions should be suitably assigned 
 to reproduce the SM fields as the zero modes. 
While a periodic boundary condition corresponding to $S^1$ 
 is taken for all of the bulk SM fields, 
 the $Z_2$ parity is assigned for gauge fields and fermions 
 in the representation ${\cal R}$ 
 by using the parity matrix $P={\rm diag}(-,-,+)$ as
\bea
A_\mu (-y) = P^\dag A_\mu(y) P, \quad A_y(-y) =- P^\dag A_\mu(y) P,  \quad 
\psi(-y) = {\cal R}(P)\psi(y) 
\label{parity}
\eea 
 where the subscripts $\mu$ ($y$) denotes the four (the fifth) 
 dimensional component. 
With this choice of parities, the $SU(3)$ gauge symmetry is 
 explicitly broken to $SU(2) \times U(1)$. 
A hypercharge is a linear combination of $U(1)$ and $U(1)'$ 
 in this setup. 
One may think that the $U(1)_X$ gauge boson which is 
 orthogonal to the hypercharge $U(1)_Y$ also has a zero mode. 
However, the $U(1)_X$ symmetry is anomalous in general 
 and broken at the cutoff scale and hence,  
 the $U(1)_X$ gauge boson has a mass of order 
 of the cutoff scale \cite{SSS}. 
As a result, zero-mode vector bosons in the model are 
 only the SM gauge fields.

Off-diagonal blocks in $A_y$ have zero modes 
 because of the overall sign in Eq.~(\ref{parity}), 
 which corresponds to an $SU(2)$ doublet. 
In fact,  the SM Higgs doublet ($H$) is identified as 
\bea
A_y^{(0)} = \frac{1}{\sqrt{2}}
\left(
\begin{array}{cc}
0 & H \\
H^\dag & 0 \\
\end{array}
\right). 
\eea
The KK modes of $A_y$ are eaten by KK modes of the SM gauge bosons 
 and enjoy their longitudinal degrees of freedom 
 like the usual Higgs mechanism.

The parity assignment also provides the SM fermions as massless modes, 
 but it also leaves exotic fermions massless. 
Such exotic fermions are made massive by introducing 
 brane localized fermions with conjugate $SU(2) \times U(1)$ charges 
 and an opposite chirality to the exotic fermions, 
 allowing us to write mass terms on the orbifold fixed points. 
In the GHU scenario, the Yukawa interaction is unified 
 with the gauge interaction, so that the SM fermions 
 obtain the mass of the order of the $W$-boson mass 
 after the electroweak symmetry breaking. 
To realize light SM fermion masses, one may introduce 
 a $Z_2$-parity odd bulk mass terms for the SM fermions, 
 except for the top quark. 
Then, zero mode fermion wave functions with opposite chirality 
 are localized towards the opposite orbifold fixed points 
 and as a result, their Yukawa coupling is exponentially 
 suppressed by the overlap integral of the wave functions. 
In this way, all exotic fermion zero modes become heavy 
 and small Yukawa couplings for light SM fermions 
 are realized by adjusting the bulk mass parameters. 
In order to realize the top quark Yukawa coupling, 
 we introduce a rank $4$ tensor representation, namely, 
 a symmetric $\overline{{\bf 15}}$ without a bulk mass \cite{CCP}. 
This leads to a group theoretical factor $2$ enhancement 
 of the top quark mass as $m_t = 2 m_W$ 
 at the compactification scale \cite{SSS}. 
Note that this mass relation is desirable 
 since the top quark pole mass receives QCD threshold 
 corrections which push up the mass about 10 GeV. 
See, for example, Ref.~\cite{FlavorCP} 
 for flavor mixing and CP violation in the GHU scenario.

\subsection{Higgs boson production through gluon fusion}
At the LHC the Higgs boson is dominantly produced  
 via gluon fusion process 
 with the following dimension 5 operator 
 between Higgs and digluon: 
\bea
{\cal L}_{{\rm eff}} = h C_{gg} G^a_{\mu\nu} G^{a\mu\nu} 
\label{dim5g}
\eea
 where $h$ is the SM Higgs boson, and $G^a_{\mu\nu}$ 
 ($a=1-8$) is the gluon field strength.   
With the setup discussed above, 
 we calculate the coefficient of this operator $C_{gg}$ 
 in our model. 
The SM contribution to $C_{gg}$ is dominated 
 by top quark 1-loop corrections. 
As a good approximation, we express the contribution 
 by using the Higgs low energy theorem \cite{HLET}, 
\bea
C_{gg}^{{\rm SM top}} \simeq \frac{g_3^2}{32\pi^2 v} 
 b_3^t \frac{\partial}{\partial \log v} \log m_t 
 = \frac{\alpha_s}{12\pi v}
\label{SMg}
\eea
 where $g_3$ ($\alpha_s=g_3^2/(4 \pi))$ is 
 the QCD coupling constant (fine structure constant), 
 $m_t$ is a top quark mass, 
 and $b_3^t=2/3$ is a top quark contribution 
 to the beta function coefficient of QCD.

In addition to the top quark contribution, 
 KK top loop contributions must be taken into account  
 in our model. 
As mentioned before, the top quark is embedded 
 into the $\overline{{\bf 15}}$-plet with a periodic boundary condition 
 and its KK mass spectrum is given by~\cite{CCP}
\bea
m_{n,t}^{(\pm)} = m_n \pm 2 m_W 
\eea
 where $m_W =80.4$ GeV is the $W$-boson mass, 
 $m_n \equiv n m_{\rm KK}$ with 
 an integer $n=1,2,3,\cdots$ and the compactification 
 scale/the unit of KK mode mass $m_{\rm KK}=1/R$. 
Although the $\overline{{\bf 15}}$-plet includes exotic massless fermions, 
 we assume that all the exotic fermions are decoupled 
 by adjusting large brane-localized mass terms 
 with the brane fermions as discussed above. 
Thus, we only consider KK modes of the SM top quark\footnote{
 One might think that the KK mode contributions from the light 
 fermions should be taken into account. 
 However, they can be safely neglected compared to those  
 from the heavy fermions since the total KK mode sum is 
 proportional to their fermion masses generated by 
 the electroweak symmetry breaking as is seen in (\ref{KKgfusion}) 
 for instance.}. 
It is straightforward to calculate KK top contributions 
 by using the Higgs low energy theorem: 
\bea
C_{gg}^{{\rm KKtop}} &\simeq& \frac{\alpha_s}{12\pi v}  
 \sum_{n=1}^\infty \frac{\partial}{\partial \log v} 
 \left[ \log (m_n + 2 m_W) + \log(m_n - 2 m_W) \right] \nonumber \\
&=& \frac{\alpha_s}{12\pi v} \sum_{n=1}^\infty 
 \left[ \frac{2 m_W}{m_n+2m_W} -\frac{2 m_W}{m_n-2m_W} \right] \nonumber \\
&\simeq&  - \frac{\alpha_s}{12\pi v} 2 \sum_{n=1}^\infty 
 \left( \frac{2 m_W}{m_n} \right)^2
=
- \frac{\alpha_s}{12\pi v} \times \frac{\pi^2}{3} \left( \frac{2 m_W}{m_{\rm KK}} \right)^2, 
\label{KKgfusion}
\eea
 where we have used an approximation $m_W^2 \ll m_n^2$
 and $\sum_{n=1}^\infty 1/n^2 = \pi^2/6$. 
Note that the KK top contribution in the gluon fusion 
 amplitude is destructive to the SM one and finite~\cite{MO}.  
This is because of the sign difference 
 between $m_{n,t}^{(+)}$ and $m_{n,t}^{(-)}$, 
 which plays a crucial role to make the KK loop 
 corrections finite\footnote{
 This finiteness is shown to be valid 
 in various space-time dimensions and at any perturbative 
 level similar to the Higgs potential in GHU \cite{Maru}.}. 
This destructive contribution is a typical feature 
 of the GHU, in sharp contrast with the one 
 in the universal extra dimension models \cite{UED}. 
Now the ratio of the Higgs production cross section 
 in our model to the SM one is estimated as 
\bea
R_\sigma \equiv 
 \left( 1 + \frac{C_{gg}^{\rm KKtop}}{C_{gg}^{\rm SMtop}} \right)^2 
\simeq \left( 1 - \frac{\pi^2}{3} \left(\frac{2 m_W}{m_{\rm KK}}\right)^2
 \right)^2. 
\label{R_g}
\eea

\subsection{Higgs decay to diphoton}
Next we calculate the KK model contributions 
 to the Higgs-to-diphoton coupling of the dimension 5 operator, 
\bea
{\cal L}_{{\rm eff}} = h C_{\gamma\gamma} F_{\mu\nu} F^{\mu\nu},  
\eea
 where $F_{\mu\nu}$ denotes the photon field strength. 
The coefficient can also be extracted from the 1-loop RGE 
 of the QED gauge coupling by using the Higgs low energy theorem. 
The diphoton coupling is induced by two contributions via 
 top quark loop and $W$-boson loop corrections. 

\subsubsection{Top quark loop contributions}
As a good approximation, top loop contribution 
 is calculated by
\bea
C_{\gamma\gamma}^{{\rm SMtop}} \simeq 
 \frac{e^2 b_1^t}{24\pi^2 v} \frac{\partial}{\partial \log v} \log m_t 
= \frac{2\alpha_{em}}{9\pi v}, 
\label{SMt2gamma}
\eea
 where $b_1=(2/3)^2 \times 3 =4/3$ is a top quark contribution 
 to the QED beta function coefficient, 
 and $\alpha_{em}$ is the fine structure constant. 
Corresponding KK top quark contribution 
 is given by
\bea
C_{\gamma\gamma}^{{\rm KKtop}} &\simeq& \frac{e^2 b_1^t}{24\pi^2 v} 
\sum_{n=1}^\infty \frac{\partial}{\partial \log v} \left[ \log (m_n + m_t)  + \log (m_n - m_t) \right] \nonumber \\
&\simeq& 
-\frac{2\alpha_{em}}{9\pi v} \times \frac{\pi^2}{3} 
 \left(\frac{2 m_W}{m_{\rm KK}} \right)^2. 
\label{KKt2gamma}
\eea
As in the case of the Higgs-to-digluon coupling, 
 the KK top contribution is destructive 
 to the SM top contribution.

\subsubsection{$W$-boson loop contributions}
The SM $W$-boson loop contribution is calculated as
\bea
C_{\gamma\gamma}^W &\simeq& \frac{e^2}{32\pi^2 v} b_1^W 
 \frac{\partial}{\partial \log v} \log m_W = -\frac{7\alpha_{em}}{8\pi v}
\label{SMW2gamma}
\eea
 where $m_W=g_2 v/2$, and $b_1^W=-7$ is a $W$-boson contribution 
 to the QED beta function coefficient. 
This is a rough estimation of the $W$-boson loop contributions 
 since $4 m_W^2/m_h^2 \gg 1$ is not well satisfied. 
For our numerical analysis in the following, 
 we actually use known loop functions 
 for the top quark and $W$-boson loop corrections.

In our model, the KK mode mass spectrum of the $W$-boson 
 is given by 
\bea
 m_{n,W}^{(\pm)} = m_n \pm m_W,  
\eea  
 so that the contribution from KK $W$-boson loop diagrams
 is found to be 
\bea
C_{\gamma\gamma}^{{\rm KKW}} &=& 
 \frac{e^2}{32\pi^2 v} b_1^W \sum_{n=1}^\infty 
 \frac{\partial}{\partial \log v} 
\left[ \log (m_n + m_W) + \log (m_n - m_W) \right]  \nonumber \\
&\simeq& 
\frac{7\alpha_{em}}{8 \pi v} \frac{\pi^2}{3} 
 \left( \frac{m_W}{m_{\rm KK}} \right)^2. 
\label{KKW2gamma}
\eea
Note again that the KK $W$-boson contribution 
 is destructive to the SM $W$-boson contribution.

Combining these results, we have 
\bea
C_{\gamma\gamma}^{{\rm SMtop}} + C_{\gamma\gamma}^W 
 &\simeq & -\frac{47\alpha_{em}}{72\pi v}, \\
C_{\gamma\gamma}^{{\rm KKtop}} + C_{\gamma\gamma}^{{\rm KKW}} 
 &\simeq& -\frac{\alpha_{em}}{216\pi v} 
 \pi^2 \left(\frac{m_W}{m_{\rm KK}} \right)^2. 
\eea
Interestingly, there is an accidental cancellation 
 between KK top and KK $W$-boson contributions. 
We find the partial decay width of $h \to \gamma \gamma$ 
 of our model to the SM one as 
\bea
R_{\gamma\gamma} \equiv \left( 1+ \frac{C_{\gamma\gamma}^{{\rm KKtop}} 
+ C_{\gamma\gamma}^{{\rm KKW}}}{C_{\gamma\gamma}^{{\rm SMtop}} + C_{\gamma\gamma}^W} \right)^2
\simeq \left( 
 1 + \frac{\pi^2}{141} \left(\frac{m_W}{m_{\rm KK}} \right)^2 
 \right)^2.  
\eea
Because of the accidental cancellation, 
 $R_{\gamma \gamma}$ is very close to one 
 for, say, $m_{\rm KK} \gtrsim  1$ TeV.

\subsection{$gg \to h \to \gamma\gamma$}
\begin{figure}[htbp]
  \begin{center}
   \includegraphics[width=80mm]{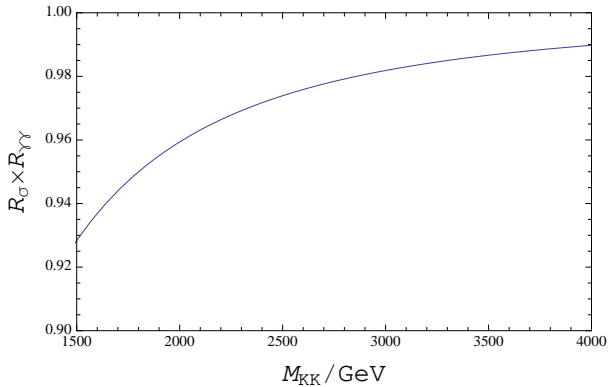}
      \end{center}
  \caption{
 The ratio of diphoton events in the simple GHU model  
 to those in the SM as a function of the compactification scale.}
\label{diphotonSM}
\end{figure}

Let us now estimate the ratio of 
 the signal strength of the process 
 $gg \to h \to \gamma \gamma$ in our model 
 to the one in the SM. 
Putting all together, we find 
\bea
{\rm R} \equiv \frac{\sigma(gg \to h \to \gamma \gamma)
}{\sigma(gg \to h \to \gamma \gamma)_{\rm SM}}
=R_\sigma \times R_{\gamma\gamma} 
\simeq 1 - \frac{374}{141} \pi^2  \left(\frac{m_W}{m_{\rm KK}} \right)^2.  
\eea
The result (using loop functions for the SM top and $W$-boson loop
 corrections) is depicted in Fig.~\ref{diphotonSM} 
 as a function of the KK mode mass/the compactification scale. 
The ratio R is found to be smaller than one, 
 because of the destructive KK mode contribution 
 to the gluon fusion channel and the accidental 
 cancellation among the KK mode contributions to 
 the Higgs-to-diphoton decay width. 
This fact has already been advocated in the previous paper 
 by the present authors \cite{MO}.

Now we extend the present GHU model to account 
 for the signal strength measured by ATLAS and CMS 
 for the process $gg \to h \to \gamma \gamma$ 
 which is considerably larger than the SM expectation. 
The simplest extension is to introduce color-singlet bulk 
 fermions with the half-periodic boundary condition,  
 $\psi(y+2\pi R) = - \psi(y)$, in the bulk. 
The main reasons for this strategy are two folds. 
The first is that since the KK mode fermion contribution 
 is destructive to the SM fermion contribution, 
 colored KK mode contribution is not desirable 
 for the Higgs production process via gluon fusion. 
On the other hand, the KK mode contribution is 
 constructive to the SM one for the Higgs-to-diphoton couple. 
Thus, the introduction of color-singlet bulk fermions 
 nicely work to enhance the diphoton signal strength. 
The second is that the half-periodic bulk fermion has 
 no massless mode, and unwanted exotic massless fermions 
 do not come out in the model. 
Another advantage of the half-periodic bulk fermion is 
 that its first KK mode mass is smaller than the compactification 
 scale and its loop corrections dominate over 
 those from the KK modes of the periodic bulk fermion. 
Furthermore, the existence of the half-periodic bulk fermion 
 is crucial to achieve a Higgs boson mass around 125 GeV 
 in our GHU model, as we will discuss in the next section.

In this paper, we consider two examples for 
 the color-singlet bulk fermions of the representations  
 ${\bf 10}$ and ${\bf 15}$ of $SU(3)$, 
 with a suitable $U(1)'$ charge assignment. 
The ${\bf 10}$-plet of $SU(3)$ is decomposed into 
 representations under $SU(2) \times U(1)$ as 
\bea
 {\bf 10} = {\bf 1}_{-1} \oplus {\bf 2}_{-1/2} 
 \oplus {\bf 3}_{0} \oplus {\bf 4}_{1/2},  
\label{10deco}
\eea
 where the numbers in the subscript denote the $U(1)$ charges. 
After the electroweak symmetry breaking of the Higgs doublet 
 ($H$: ${\bf 2}_{1/2}$), the KK mass spectrum 
 is found as follows: 
\bea
&& 
\left( m_{n,-1}^{(\pm)} \right)^2 
 = \left( m_{n+\frac{1}{2}} \pm 3 m_W \right)^2 +M^2,~~ 
   \left( m_{n+\frac{1}{2}} \pm   m_W \right)^2 +M^2, \nonumber \\ 
&& 
\left( m_{n,0}^{(\pm)} \right)^2 
 = \left( m_{n+\frac{1}{2}} \pm 2 m_W \right)^2 +M^2,~~ 
   m_{n+\frac{1}{2}}^2 + M^2, \nonumber \\ 
&& 
\left( m_{n,+1}^{(\pm)} \right)^2 
 = \left( m_{n+\frac{1}{2}} \pm m_W \right)^2 +M^2, \nonumber \\ 
&& 
\left( m_{n,+2}^{(\pm)} \right)^2 
 = m_{n+\frac{1}{2}}^2 +M^2, 
\eea  
 where the numbers in the subscript denote 
 the ``electric charges"\footnote{
 Here ``electric charges" mean by electric charges 
 of $SU(2) \times U(1) \supset SU(3)$. 
 A true electric charge of each KK mode is given 
 by a sum of the ``electric charge" and $U(1)'$ charge $Q$.
 }  
 of the corresponding KK mode  fermions, 
 $m_{n+\frac{1}{2}}= \left( n+\frac{1}{2}\right) m_{\rm KK}$ 
 with $n=0,1,2,\cdots$, and $M$ is a bulk mass.

Employing the Higgs low energy theorem with these mass spectrum, 
 we calculate the ${\bf 10}$-plet KK mode contributions 
 to the Higgs-to-diphoton coupling as 
\bea
C_{\gamma\gamma}^{{\rm KK-10}} 
 \simeq (Q-1)^2 F(3m_W) + (Q-1)^2 F(m_W) + Q^2 F(2m_W) + (Q+1)^2 F(m_W) 
\label{lepton10}
\eea
 where $Q$ is a $U(1)^\prime$ charge for the ${\bf 10}$-plet, 
 and the function $F(m_W)$ is defined as 
\bea
F(m_W) &\equiv& \frac{\alpha_{em}}{6\pi v} 
 \sum_{n=1}^\infty \frac{\partial}{\partial \log v} \left[ \log \sqrt{M^2+(m_{n + \frac{1}{2}} + m_W)^2} 
+ \log \sqrt{M^2+(m_{n + \frac{1}{2}} - m_W)^2} \right] \nonumber \\
&=& \frac{\alpha_{em}}{6\pi v} m_W 
\sum_{n=0}^\infty 
\left( 
\frac{ m_{n+\frac{1}{2}}+m_W }{(m_{n+\frac{1}{2}}+m_W)^2+M^2} 
+
\frac{ m_{n+\frac{1}{2}}-m_W }{(m_{n+\frac{1}{2}}-m_W)^2+M^2} 
\right) \nonumber \\
&\simeq&
-\frac{\alpha_{em}}{3\pi v} \left(\frac{m_W}{m_{\rm KK}} \right)^2
\sum_{n=0}^\infty 
\frac{\left( n+\frac{1}{2} \right)^2- c_B^2}
{\left( \left(n+\frac{1}{2}\right)^2+ c_B^2 \right)^2} \nonumber \\ 
&=&
-\frac{\alpha_{em}}{6 \pi v} \left(\frac{m_W}{m_{\rm KK}} \right)^2
 \frac{\pi^2}{\cosh (\pi c_B)}. 
\eea
Here we have used the approximation $m_W^2 \ll m_{\rm KK}^2$, 
 and $c_B \equiv M/m_{\rm KK}$.

For the ${\bf 15}$-plet case, 
 the decomposition under $SU(2) \times U(1)$ is given as 
\bea
 {\bf 15} = {\bf 1}_{-4/3} \oplus {\bf 2}_{-5/6} 
 \oplus {\bf 3}_{-1/3} \oplus {\bf 4}_{1/6} 
 \oplus {\bf 5}_{2/3}.  
\label{15deco}
\eea
After the electroweak symmetry breaking, 
 the KK mass spectrum is found as follows: 
\bea
&& 
\left( m_{n,-4/3}^{(\pm)} \right)^2 
 = \left( m_{n+\frac{1}{2}} \pm 4 m_W \right)^2 +M^2,~~ 
   \left( m_{n+\frac{1}{2}} \pm 2  m_W \right)^2 +M^2,~~  
   m_{n+\frac{1}{2}}^2+M^2, 
  \nonumber \\ 
&& 
\left( m_{n,-1/3}^{(\pm)} \right)^2 
 = \left( m_{n+\frac{1}{2}} \pm 3 m_W \right)^2 +M^2,~~~ 
   \left( m_{n+\frac{1}{2}} \pm   m_W \right)^2 +M^2, 
  \nonumber \\ 
&& 
\left( m_{n,2/3}^{(\pm)} \right)^2 
 = \left( m_{n+\frac{1}{2}} \pm 2 m_W \right)^2 +M^2,~~ 
   m_{n+\frac{1}{2}}^2+M^2, 
  \nonumber \\ 
&&
\left( m_{n,5/3}^{(\pm)} \right)^2 
 = \left( m_{n+\frac{1}{2}} \pm m_W \right)^2 +M^2,
  \nonumber \\ 
&&
\left( m_{n,8/3}^{(\pm)} \right)^2 
 = m_{n+\frac{1}{2}}^2+M^2, 
\eea  
 where the numbers in the subscript denote the ``electric charges"  
 of the corresponding KK fermions. 
In this case, the Higgs-to-diphoton coupling is calculated as 
\bea
C_{\gamma\gamma}^{{\rm KK-15}} &\simeq& 
 (Q-4/3)^2 F(4m_W) + (Q-4/3)^2 F(2m_W) \nonumber \\
&+& (Q-1/3)^2 F(3m_W) + (Q-1/3)^2 F(m_W) \nonumber \\
&+& (Q+2/3)^2 F(2m_W) + (Q+5/3)^2 F(m_W).
\label{lepton15}
\eea

\begin{figure}[htbp]
  \begin{center}
   \includegraphics[width=70mm]{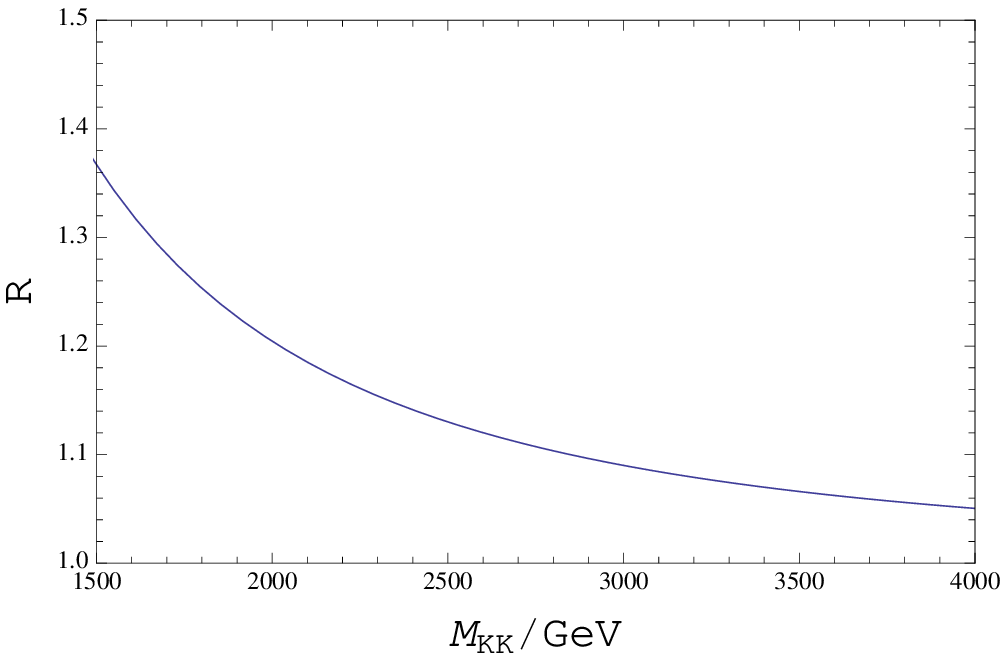}
   \hspace*{10mm}
   \includegraphics[width=70mm]{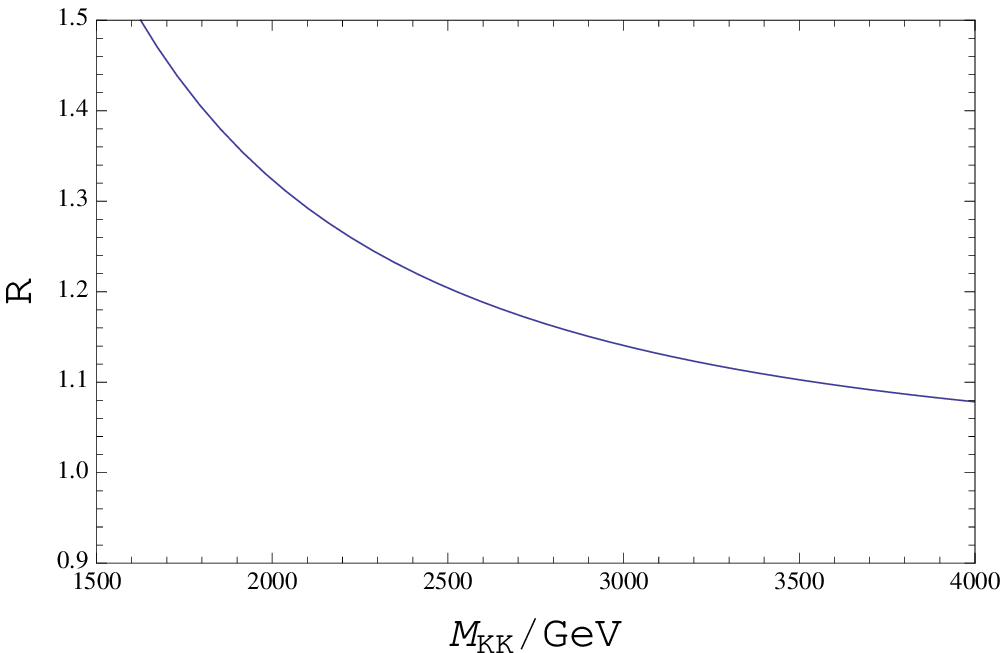}
   \end{center}
  \caption{
The diphoton signal strength (normalized by the SM prediction) 
 in the GHU model with the ${\bf 10}$-plet (left) 
 and ${\bf 15}$-plet (right) bulk fermions
 as a function of the compactification scale.
Here we have used $Q=-1$ and $c_B=0.23$ for 
 the left panel, while 
 $Q=-5$ and $c_B=0.69$ for the right panel.
}
  \label{diphotonwithleptonR}
\end{figure}

For the two cases, we plot the ratio R as a function of  
 the KK mode mass $m_{\rm KK}$ in Fig.~\ref{diphotonwithleptonR}. 
The left panel corresponds to the case 
 with the ${\bf 10}$-plet bulk fermion, 
 where we have fixed $Q=-1$ and $c_B=0.23$. 
As we will see in the next section, 
 the Higgs boson mass around 125 GeV
 can be reproduced with the bulk mass $c_B=0.23$ 
 for $m_{\rm KK}=3$ TeV. 
The result for the case with the ${\bf 15}$-plet bulk fermion 
 is depicted in the right panel for $Q=-5$ and $c_B=0.69$. 
This bulk mass reproduces the Higgs boson mass around 125 GeV. 
We find the Higgs-to-diphoton signal strength is 
 considerably enhanced in the presence of 
 the half-periodic bulk fermions with the TeV scale mass.

\begin{figure}[htbp]
  \begin{center}
   \includegraphics[width=70mm]{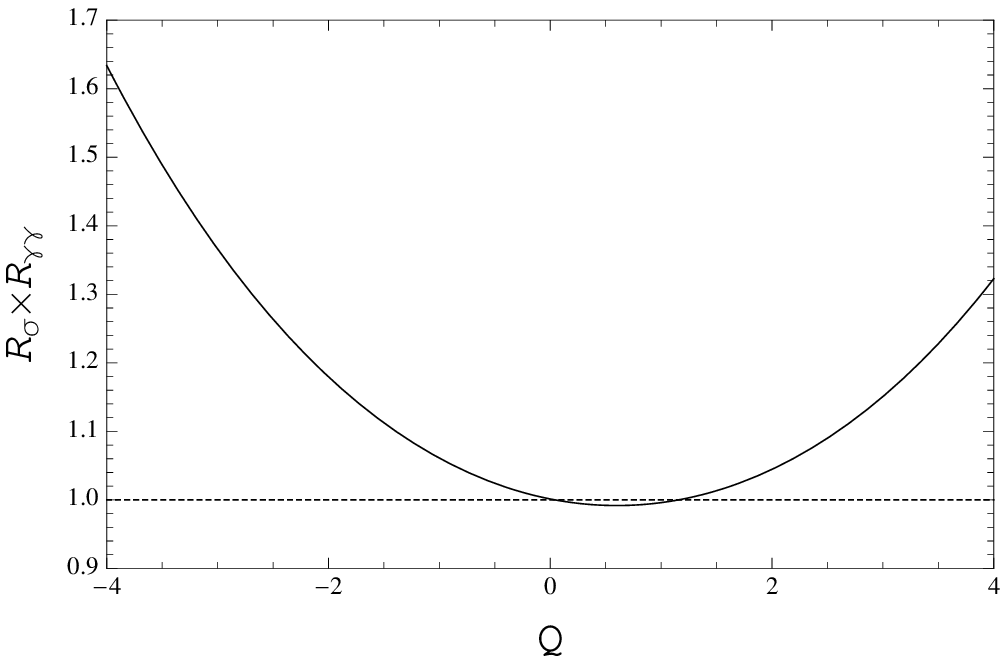}
   \hspace*{10mm}
   \includegraphics[width=70mm]{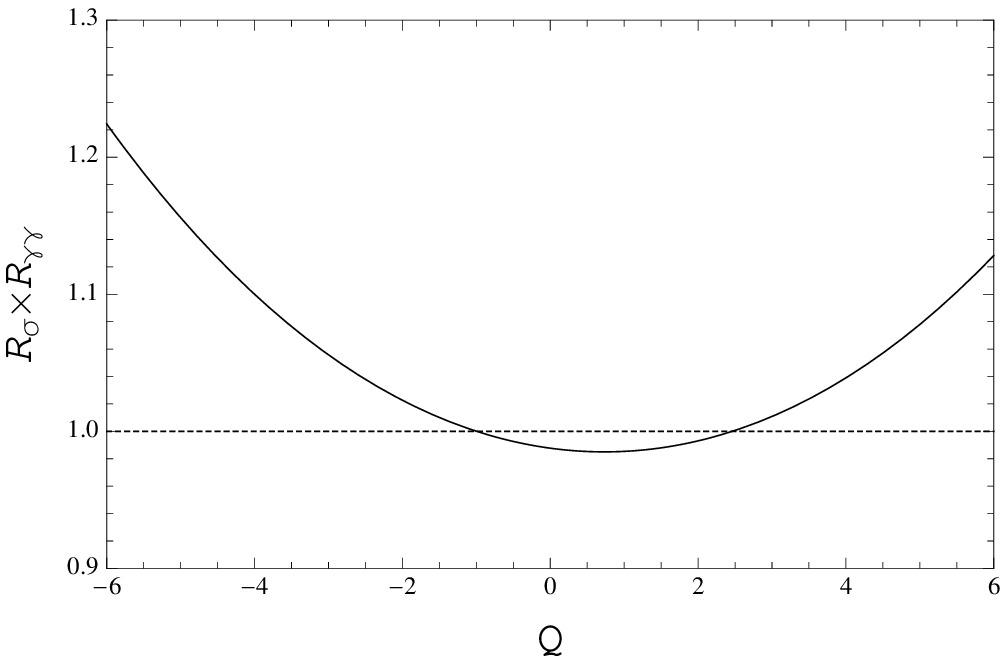}
   \end{center}
  \caption{
The diphoton signal strength (normalized by the SM prediction) 
 in the GHU model with the ${\bf 10}$-plet (left) 
 and ${\bf 15}$-plet (right) bulk fermions
 as a function of the $U(1)'$ charge $Q$, 
 for $m_{\rm KK}=3$ TeV. 
The bulk masses are fixed as $c_B=0.23$ for 
 the left panel, while $c_B=0.69$ for the right panel.
}
 \label{diphotonwithleptonQ}
\end{figure}

As can be understood from Eqs.~(\ref{lepton10}) and (\ref{lepton15}), 
 the rate of the enhancement depends on the choice 
 of $U(1)'$ charge $Q$.  
In other words, it can be large as we like by adjusting a $U(1)'$ charge. 
In Fig.~\ref{diphotonwithleptonQ}, 
 we plot the ratio of diphoton signal strength to 
 the SM one as a function of the $U(1)'$ charge $Q$, 
 for the two cases. 
For each plot, the bulk masses are fixed 
 to be the same values as in the previous plots. 
We can see that $|Q|={\cal O}(1)$ is enough 
 to give rise to an order 10\% enhancement 
 of the diphoton signal. 
In general, a larger representation field 
 leads to more enhancements than those by smaller 
 representations, since large representations 
 include more fields with higher $U(1)$ charges 
 in the SM decomposition. 
In Fig.~\ref{diphotonwithleptonQ}, 
 the deviation of the diphoton signal strength 
 for the case with the ${\bf 15}$-plet fermion  
 is milder than the case with the ${\bf 10}$-plet. 
This is because  a large bulk mass is assigned 
 for the case with the ${\bf 15}$-plet fermion
 and the KK modes are heavier.

\section{Estimate of Higgs boson mass}
In this section, we discuss how the Higgs boson mass 
 around 125 GeV is realized in our model. 
Realizing the 125 GeV Higgs boson mass as well as 
 the electroweak symmetry breaking 
 is a quite non-trivial phenomenological issue 
 in 5-dimensional GHU scenario. 
This is because the Higgs doublet is embedded 
 in the five dimensional component  of the bulk gauge  
 field and as a result, the Higgs doublet has 
 no scalar potential at the tree level. 
The electroweak symmetry should be broken dynamically, 
 in other words, at the quantum  level. 
In addition, a calculated Higgs boson mass is most likely 
 to be small, since the Higgs quartic coupling is 
 generated at loop levels. 
Towards a realistic GHU scenario, a variety of extra bulk fields 
 with suitable boundary conditions and bulk/brane mass terms 
 have been considered (see, for example, Refs.~\cite{SSS, CCP}). 
It is a highly non-trivial task to propose 
 a simple and phenomenologically viable GHU model.

In estimating Higgs boson mass, 
 we take a 4-dimensional effective theory approach 
 developed by Ref.~\cite{GHcondition}.
As has been shown in this paper, 
 the low energy effective theory of the 5-dimensional GHU scenario 
 is equivalent to the SM with the so-called ``gauge-Higgs condition'' 
 on the Higgs quartic coupling, 
 namely, we impose a vanishing Higgs quartic coupling
 at the compactification scale. 
This boundary condition reflects the 5-dimensional gauge 
 invariance which gets restored at an energy 
 higher than the compactification scale. 
Employing this effective theory approach, the Higgs boson mass 
 at low energies is easily calculated by solving the RGE 
 of the Higgs quartic coupling with the gauge-Higgs condition, 
 instead of calculating the Coleman-Weinberg potential 
 of the Higgs doublet. 
We assume that the electroweak symmetry breaking correctly occurs  
 by the introduction of a suitable set of bulk fermions. 
Note that the effective Higgs mass squared is quadratically 
 sensitive to the mass of heavy states, 
 while the effective Higgs quartic coupling is dominantly  
 determined by interactions of the Higgs doublet with light states. 
Therefore, the Higgs boson mass at low energies 
 is mainly determined by light states below the compactification scale, 
 once we assume the correct electroweak symmetry breaking.

In our model, we have introduced bulk fermions 
 with the half-periodic boundary condition, 
 and their first KK modes appear 
 below the compactification scale. 
Therefore, not only the SM particles but also 
 the first KK modes are involved in our RGE analysis 
 with the gauge-Higgs condition\footnote{
In Ref.~\cite{GHUcond-mh}, the gauge-Higgs condition
 with only the SM particle contents below the compactification scale 
 is used to predict Higgs boson mass.}.
The 1-loop RGE for the Higgs quartic coupling $\lambda$ 
 below the compactification scale is given by 
\bea
\frac{d \lambda}{d \ln \mu} &=& 
\frac{1}{16\pi^2} 
\left[
12 \lambda^2 
- \left( \frac{9}{5} g_1^2 + 9 g_2^2 \right) \lambda
+ \frac{9}{4} \left( \frac{3}{25}g_1^4 + \frac{2}{5}g_1^2 g_2^2 + g_2^4 \right) \right. \nonumber \\
&& \left. + 4 \left( 
  3 y_t^2 + C_2({\bf R}) \left( \frac{g_2}{\sqrt{2}}\right)^2 
 \right) \lambda 
 -4 \left(3 y_t^4 + C_4({\bf R})
 \left( \frac{g_2}{\sqrt{2}}\right)^4 
 \right)  \right], 
\label{RGEup}
\eea
 where $y_t$ is the top Yukawa coupling, 
 $g_{1,2}$ are the $SU(2)$, $U(1)_Y$ gauge couplings, respectively,
 and $C_2({\bf R})$  and $C_4({\bf R})$ are 
 contributions to the beta function 
 by the representation ${\bf R}={\bf 10}$ or ${\bf 15}$. 
In our RGE analysis, we neglect the KK mode mass splitting 
 by the electroweak symmetry breaking 
 and set the first KK mode mass as 
\bea 
 m_0^{(\pm)}= \frac{1}{2} m_{\rm KK} \sqrt{1+4 c_B^2}. 
\eea 
For the energy scale $m_0^{(\pm)} \leq \mu \leq m_{\rm KK}$, 
 the coefficients $C_2({\bf R})$ and $C_4({\bf R})$ 
 are explicitly given by 
\bea
C_2({\bf 10}) &=& 2 \left( 3^2 + 1^2 + 2^2 + 1^2 \right), 
\nonumber \\ 
C_4({\bf 10}) &=& 2 \left( 3^4 + 1^4 + 2^4 + 1^4 \right), 
\nonumber \\ 
C_2({\bf 15}) &=& 2 \left( 4^2 + 2^2 + 3^2 + 1^2 + 2^2 + 1^2 \right), 
\nonumber \\ 
C_4({\bf 15}) &=& 2 \left( 4^4 + 2^4 + 3^4 + 1^4 + 2^4 + 1^4 \right), 
\eea
 while these coefficients are set to be $0$ for $\mu < m_0^{(\pm)}$. 
In our analysis, the running effects 
 for $y_t, g_{1,2}$ are simply neglected.

\begin{figure}[htbp]
  \begin{center}
   \includegraphics[width=90mm]{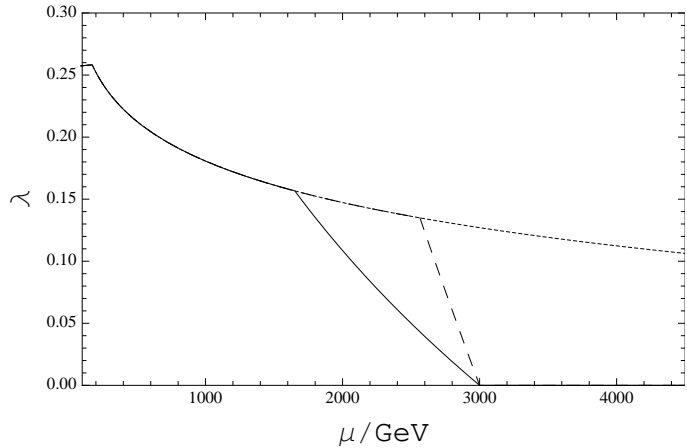}
  \end{center}
  \caption{
1-loop RGE running of the Higgs quartic coupling. 
The solid (dashed) line corresponds  to the case 
 of the ${\bf 10}$-plet (${\bf 15}$-plet) bulk fermion 
 with the bulk mass $c_B=0.23$ ($c_B=0.69$). 
Here the compactification scale is fixed 
 as $m_{\rm KK}=1/R = 3$ TeV, at which 
 the gauge-Higgs condition ($\lambda(m_{\rm KK})=0$) 
 is applied. 
The dotted line shows the running of 
 the SM Higgs quartic coupling 
 with the boundary condition,
 $\lambda(\mu=m_h)=0.258$, corresponding 
 to the Higgs pole mass $m_h=125$ GeV. 
}
  \label{RGElambda}
\end{figure}

The numerical results of 1-loop RGE of the Higgs quartic coupling 
 are shown in Fig.~\ref{RGElambda}. 
Here we have applied the gauge-Higgs condition 
 at the compactification scale $1/R = m_{\rm KK}=3$ TeV 
 and numerically solve the RGE toward low energies. 
In the analysis, we have used 
 $y_t(\mu) =0.943$ for $\mu \geq m_t=173.1$ GeV 
 ($y_t(\mu)=0$ for $\mu < m_t=173.1$ GeV), 
 and $g_1=0.459$, and $g_2=0.649$ at the $Z$-boson mass scale. 
For simplicity, we estimate the Higgs boson pole mass 
 by the condition $ \lambda(\mu=m_h) v^2 =m_h^2$. 
In Fig.~\ref{RGElambda}, the bulk masses of 
 the ${\bf 10}$-plet and the ${\bf 15}$-plet 
 are fixed to be the values used in the previous section, 
 $c_B=0.23$ and $ c_B=0.69$, respectively, 
 with which Higgs boson mass of $m_h=125$ GeV 
 (equivalently, $\lambda (\mu=m_h) = 0.258$) is realized. 
The solid (dashed) line represents the running Higgs quartic 
 coupling for the case with the ${\bf 10}$-plet (${\bf 15}$-plet) 
 bulk fermion, 
 while the dotted line corresponds to the RGE running 
 in the SM case with the boundary condition  
 $\lambda (\mu=m_h) = 0.258$. 
In this rough analysis, the Higgs quartic coupling becomes zero 
 at $\mu \sim 10^{4.5}$ GeV. 
As is well-known, in more precise analysis with higher 
 order corrections (see, for example, \cite{RGE-Higgs}), 
 the Higgs quartic coupling becomes zero 
 at $\mu \sim 10^{10}$ GeV. 
In the precise analysis, the running top Yukawa coupling 
 is monotonically decreasing and the higher order corrections 
 positively contribute to the beta function, 
 as a result, the scale realizing $\lambda(\mu)=0$ 
 is pushed up to high energies.

As can be seen from Fig.~\ref{RGElambda},
 the existence of the half-periodic bulk fermions 
 is essential to realize the Higgs mass around 125 GeV 
 with the compactification at the TeV scale. 
Since the bulk fermions provide many first KK mode 
 fermions in the SM decomposition, 
 the running Higgs quartic coupling is sharply 
 rising from zero toward low energies. 
In addition, the bulk mass also plays a crucial role 
 to adjust the resultant Higgs boson mass to be 125 GeV.

\section{Conclusion}
In this paper, we have revisited the Higgs boson production 
 via gluon fusion and its diphoton decay process 
 in a 5-dimensional $SU(3)\times U(1)'$ GHU model. 
As shown in \cite{MO} and re-confirmed in this paper, 
 the diphoton signal events is reduced 
 in a simple GHU model with only the KK modes 
 of the SM top quark and $W$-boson 
 taken into account. 
As a simple extension of the simple model, 
 we have introduced color-singlet bulk fermions 
 with the half-periodic boundary condition and bulk masses. 
For concreteness, we have considered the $SU(3)$ 
 ${\bf 10}$-plet and the ${\bf 15}$-plet 
 with a $U(1)^\prime$ charge $Q$. 
With the charge $Q$ being of order one, 
 the diphoton signal events can be remarkably enhanced 
 from the SM prediction by 1-loop corrections of the KK modes  
 at the TeV scale. 
The signal significance of the diphoton events 
 observed at ATLAS and CMS shows a positive deviation form the SM 
 expectation and it can be an indirect signal of the GHU model. 
The bulk fermions also play a crucial role to yield 
 the observed Higgs boson mass around 125 GeV. 
Employing the gauge-Higgs condition, we have shown 
 in the RGE analysis that a Higgs boson mass, 
 which is predicted to be too small in the simple GHU model, 
 is dramatically enhanced in the presence of 
 the half-periodic bulk fermions and the 125 GeV Higgs 
 boson mass is reproduced by adjusting the fermion bulk masses.

There was an interesting argument of naturalness 
 in Ref.~\cite{AHBDAF} that if a large diphoton signal excess 
 is caused by extra fermions having strong couplings with the Higgs boson, 
 the Higgs potential becomes unstable far beneath the 10 TeV scale. 
In order to keep the Higgs potential stable up to 
 a very high energy, we need a new light scalar field 
 with a strong coupling to the Higgs doublet, 
 which positively contributes to the beta function 
 of the Higgs quartic coupling. 
One may think that the existence of such a new light 
 scalar field makes the Standard Model more unnatural~\cite{AHBDAF}. 
However, note that in the 5-dimensional GHU model, the vanishing of 
 the running Higgs quartic coupling indicates not 
 the instability of the Higgs potential but instead, 
 the restoration of the bulk gauge symmetry 
 which should occur at an energy higher than 
 the compactification scale. 
Therefore, the GHU scenario offers a natural solution 
 to the instability of the  Higgs potential.

There is another interesting feature of our model. 
As discussed in \cite{GHDM} (see also \cite{GHDM2}), 
 the lightest KK mode of the half-periodic bulk fermion, 
 independently of the background metric, is stable 
 in the effective 4-dimensional theory due to the remaining 
 accidental discrete symmetry. 
Thus, if its electric charge is arranged to be 0, 
 the lightest KK mode becomes a good candidate 
 for the dark matter in the present Universe. 
For detailed discussion on the dark matter physics 
 in the GHU model, we refer Ref.~\cite{GHDM}.

One may be also interested in the KK mode contributions 
 to the $h \to Z \gamma$ process and its correlation 
 to the $h \to \gamma\gamma$ process. 
However, it seems that the analysis is not so straightforward 
 since the Higgs low energy theorem for $h \to Z \gamma$ 
 is not very clear and the corresponding loop functions 
 are complicated. 
We leave this subject for the future study.

\section*{Acknowledgments}
The work of N.M. is supported in part by the Grant-in-Aid 
 for Scientific Research from the Ministry of Education, 
 Science and Culture, Japan No. 24540283 
  and by Keio Gijuku Academic Development Funds.
The work of N.O. is supported in part 
 by the DOE Grant No. DE-FG02-10ER41714.


\end{document}